\documentclass[fleqn,useAMS]{Papers}
\usepackage{newtxtext,newtxmath}
\usepackage[T1]{fontenc}
\usepackage{subfloat}
\usepackage{makecell}

\usepackage[square,sort,numbers]{natbib}
\usepackage{graphicx}	% Including figure files
\usepackage{amsmath}	% Advanced maths commands
\usepackage{lipsum}
\usepackage{caption}
\usepackage{comment}
\usepackage{textgreek} 
\usepackage{xcolor}  
\newcommand{\rev}[1]{{\color{black} #1}} 

\title[Maximum entropy distributions of dark matter]{Maximum entropy distributions of dark matter in \texorpdfstring{$\Lambda$}{}CDM cosmology}
\author[Z. Xu]{Zhijie (Jay) Xu,$^{1}$\thanks{E-mail: \href{mailto:zhijie.xu@pnnl.gov}{zhijie.xu@pnnl.gov}; \href{mailto:zhijiexu@hotmail.com}{zhijiexu@hotmail.com}}
\\
$^{1}$Physical and Computational Sciences Directorate, Pacific Northwest National Laboratory; Richland, WA 99354, USA\\
}

\date{Accepted XXX. Received YYY; in original form ZZZ}

\pubyear{2022}

\begin{document}
\label{firstpage}
\pagerange{\pageref{firstpage}--\pageref{lastpage}}
\maketitle

\begin{abstract}
Small-scale challenges to $\Lambda$CDM cosmology require a deeper understanding of dark matter physics.
This paper aims to develop the maximum entropy distributions for dark matter particle velocity (denoted by $X$), speed (denoted by $Z$), and energy (denoted by $E$) that are especially relevant on small scales where system approaches full virialization.
For systems involving long-range interactions, \rev{a spectrum of halos of different sizes is required to form to maximize system entropy.} While the velocity in halos can be Gaussian, the velocity distribution throughout the entire system, involving all halos of different sizes, is non-Gaussian. With the virial theorem for mechanical equilibrium, we applied the maximum entropy principle  to the statistical equilibrium of entire system, such that the maximum entropy distribution of velocity (the $X$ distribution) could be analytically derived. The halo mass function was not required in this formulation, but it did indeed result from the maximum entropy.
The predicted $X$ distribution involves a shape parameter $\alpha$ and a velocity scale, $v_0$. The shape parameter $\alpha$ reflects the nature of force ($\alpha\rightarrow0$ for long-range force or $\alpha\rightarrow\infty$ for short-range force). Therefore, the distribution approaches Laplacian with $\alpha\rightarrow0$ and Gaussian with $\alpha\rightarrow\infty$. For an intermediate value of $\alpha$, the distribution naturally exhibits a Gaussian core for $v\ll v_0$ and exponential wings for $v\gg v_0$, as confirmed by N-body simulations. From this distribution, the mean particle energy of all dark matter particles with a given speed, $v,$ follows a parabolic scaling for low speeds ($\propto v^2$ for $v\ll v_0$ in halo core region, i.e., "Newtonian") and a linear scaling for high speeds ($\propto v$ for $v\gg v_0$ in halo outskirt, i.e., exhibiting "non-Newtonian" behavior in MOND due to long-range gravity). We compared our results  against N-body simulations and found a good agreement.
\end{abstract}

\begin{keywords}
\vspace*{-10pt}
Dark matter; Maximum entropy distributions; Theoretical models; Small scale challenges
\end{keywords}

%%%%%%%%%%%%%%%%%%%%%%%%%%%%%%%%%%%%%%%%%%%%%%%%%
%%%%%%%%%%%%%%%%% BODY OF PAPER %%%%%%%%%%%%%%%%%%
\begingroup
\let\clearpage\relax
\tableofcontents
\endgroup
\vspace*{-20pt}
%\newpage

\section{Introduction}
\label{sec:1}
Over the last forty years, the standard $\Lambda$CDM model has significantly enhanced our understanding of large-scale structures, the state of early universe, and the content of different kinds of matter and energy \citep{Peebles:1984-Tests-of-cosmological-models,Peebles:2012-Seeing-Cosmology-Grow, Spergel:2003-First-Year-Wilkinson-Microwave-Anisotropy,Komatsu:Seven-year-Wilkinson-Microwave-Anisotropy-Probe,Frenk:2012-Dark-matter-and-cosmic-structure}. However, the theory has encountered a number of challenges on smaller scales \citep{Bullock:2017-Small-Scale-Challenges-to-the,DelPopolo:2017-Small-scale-problems-of-the,Perivolaropoulos:2022-Challenges-for}, where the model predictions are shown to be inconsistent with observations. The main challenges include: the cusp-core problem \citep{Flores:1994-Observational-and-Theoretical-Constraints,Moore:1994-Evidence-against-dissipation-less-dark-matter,deBlok:2009-The-Core-Cusp-Problem}, missing satellite problem \citep{Klypin:1999-Where-Are-the-Missing-Galactic,Moore:1999-Dark-Matter-Substructure}, "too-big-to-fail" problem \citep{Boylan_Kolchin:2011-Too-big-to-fail,Boylan-Kolchin:2012-The-Milky-Ways-bright-satellites}. In particular, the physics behind the tight baryonic Tully-Fisher relation and MOND (modified Newtonian dynamics) still remains unclear within the $\Lambda$CDM framework \citep{McGaugh:2000-The-baryonic-Tully-Fisher-rela,Famaey:2013-Challenges-fo-CDM-and-MOND}. These small-scale challenges require a better understanding of baryon and dark matter physics on small scales (<1 Mpc), either within or beyond the $\Lambda$CDM cosmology. 

One important aspect may be related to dark matter (DM) velocity distributions, which are also critical for the direct and indirect dark matter searches. For hydrodynamic turbulence, the velocity distribution is almost Gaussian and in equilibrium on a large scale or there is a Maxwell-Boltzmann distribution in speed. This is a direct result of maximizing entropy for systems involving short-range interactions, as we have learned from the kinetic theory of gases. With kinetic energy cascading down to smaller scales, the distribution of velocity deviates from Gaussian and becomes more and more skewed, with the increasing skewness due to the viscous interactions on small scales. 

The situation is quite different for the flow of dark matter, where the gravitational interaction is long-ranging and kinetic energy is cascading from small to large scales \citep{Xu:2023-Universal-scaling-laws-and-density-slope}. 
Mass and energy cascades provide non-equilibrium dark matter flow a mechanism that continuously releases energy and maximizes system entropy \citep{Xu:2021-Inverse-mass-cascade-mass-function}. Unlike the hydrodynamic turbulence, the collisionless dark matter flow should approach virial equilibrium on small scales, such that the DM velocity distribution on small scale should also approach some maximum entropy distribution (denoted by $X$). This paper aims to find this distribution for the flow of dark matter. With kinetic energy cascaded to larger scales, DM velocity distribution gradually deviates from the $X$ distribution with increasing skewness \citep{Xu:2022-Two-thirds-law-for-pairwise-ve}.  

The maximum entropy distributions of dark matter in $\Lambda$CDM cosmology concern the final stationary state after relaxation (a limiting state that can never reach). These distributions are especially relevant on small scales where system approaches full virialization. The original problem was motivated by the paradox between apparent universally stable self-gravitating structures and the extremely long, unphysical, two-body relaxation time required to form these structures. Ogorodnikov \citep{Ogorodnikov:1957-Statistical-Mechanics-of-the-Simplest} and Lynden-Bell \citep{Lyndenbell:1967-Statistical-Mechanics-of-Viole} were among the first to seek a fast relaxation mechanism to drive system toward the final equilibrium. The process of "violent relaxation" was thus introduced \citep{Lyndenbell:1967-Statistical-Mechanics-of-Viole} to describe the fast energy exchange between the rapid fluctuation of gravitational potential and collisionless particles moving through the potential field. In the same paper, statistical mechanics subject to an exclusion principle was developed, where two parcels of phase space are precluded from superimposing. The theory predicts an isothermal sphere and Maxwellian velocity distribution as the final equilibrium state with maximum entropy. 

However, that prediction is not entirely satisfactory. The predicted isothermal spheres have infinite mass even though prediction was made with apparent constraints of fixed finite energy and mass. The large scale N-body simulations \citep{Navarro:1995-Simulations-of-X-Ray-Clusters,Navarro:1997-A-universal-density-profile-fr,Einasto:1989-Galactic-Models-with-Massive-C,Merritt:2006-Empirical-models-for-dark-matt} for structure formation have revealed a remarkably universal but non-isothermal halo density profile that cannot be explained by that theory. Therefore, to better understand small-scale challenges, a new theory on the limiting distributions of dark matter at small scales is desirable. 

Despite significant progress made over the past several decades \citep{Shu:1978-Statistical-Mechanics-of-Viole,Tremaine:1986-H-Functions-and-Mixing-in-Viol,White:1987-Maximum-Entropy-States-and-the,Hjorth:2010-Statistical-Mechanics-of-Colli,Kull:1997-Note-on-the-statistical-mechan,Williams:2010-Statistical-Mechanics-of-Collisionless-Orbits}, the statistical mechanics of a self-gravitating collisionless system remains a long-standing puzzle. The difficulty can be partially attributed to the unshielded, long-range gravitational force, associated negative heat capacity, and lack of equivalence between canonical and microcanonical ensembles \citep{Padmanabhan:1990-Statistical-Mechanics-of-Gravi}. In contrast, the collisional molecular gases have short-range interactions and plasma systems have an effective short-range interaction due to Debye shielding. Because of these fundamental differences, conventional statistical mechanics for systems characterized by short-range interactions  cannot be directly applied. Hence, it is also necessary to develop new theory that can handle long-range interactions, from which maximum entropy distributions of entire system can be obtained.

\section{Limiting probability distributions}
\label{sec:2}
\subsection{Statement of the problem}
\label{sec:2.1}
We considered a system of \textit{N} particles interacting through a two-body power-law potential $V\left(r\right)$ with an arbitrary exponent \textit{n}, namely, $V\left(r\right)\propto r^{n} $. In particular, the case $n=-1$ represents the usual gravitational interaction. The spatial distribution of collisionless dark matter consists of distinct clusters (halos) of different sizes \citep{Neyman:1952-A-Theory-of-the-Spatial-Distri, Cooray:2002-Halo-models-of-large-scale-str, Jenkins:2001-The-mass-function-of-dark-matt,Colberg:1999-Linking-cluster-formation-to-l,Moore:1999-Cold-collapse-and-the-core-cat}. We will demonstrate that the spatial distribution of dark matter is dependent on the potential exponent, $n$. The formation of halo structures is essentially an intrinsic feature to maximize entropy for systems involving long-range interactions. 

\begin{figure}
\includegraphics*[width=\columnwidth]{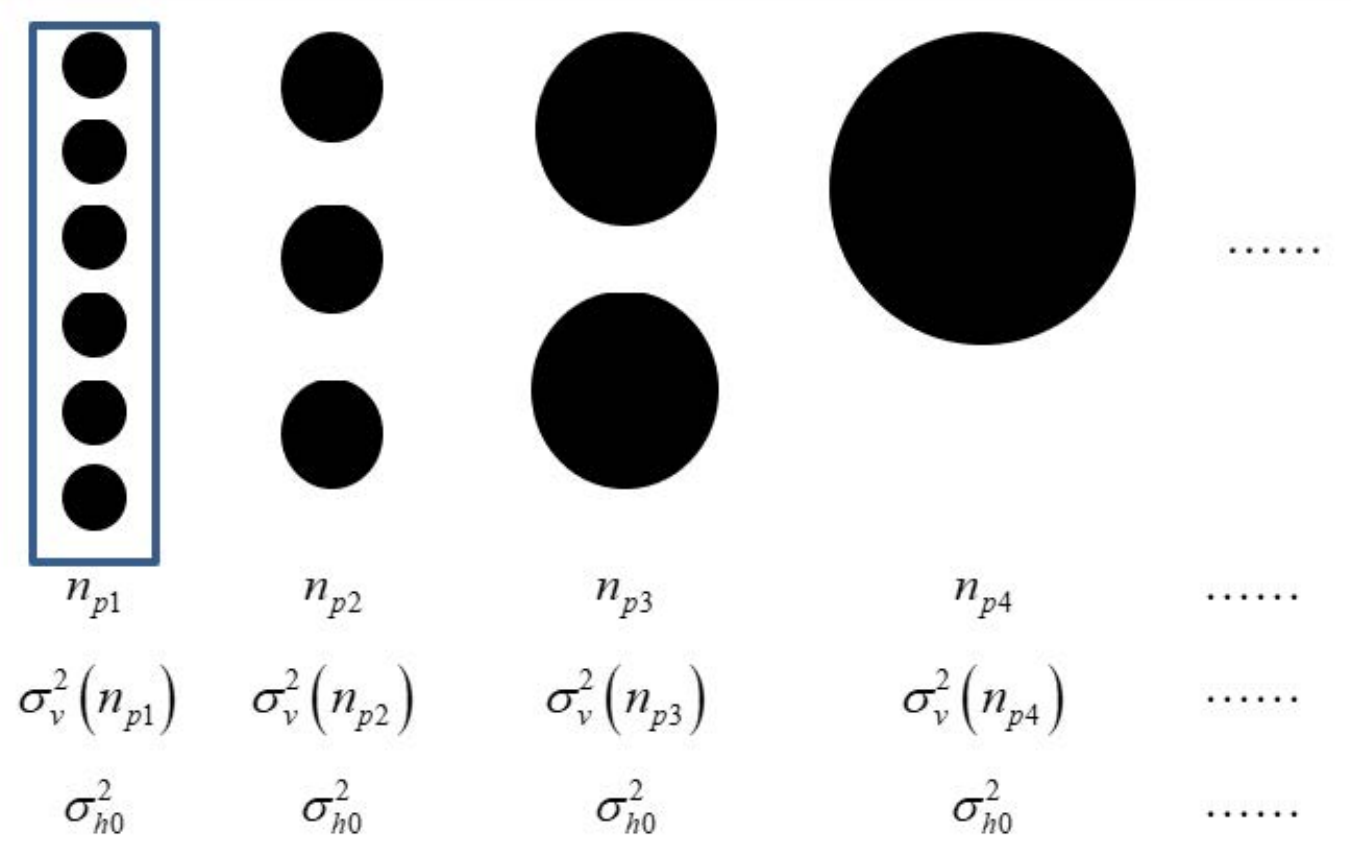}
\caption{Schematic plot of halo groups of different sizes. Halos are grouped and sorted according to the number of particles $n_{p}$ in the halo, with increasing size shown from left to right. Every group of halos of the same size, $n_{p} $, is characterized by a halo virial dispersion $\sigma _{v}^{2} \left(n_{p} \right)$ as a function of halo size, while the halo velocity dispersion, $\sigma _{h}^{2} =\sigma _{h0}^{2}$, is relatively independent of halo size.}
\label{fig:1}
\end{figure}

Figure \ref{fig:1} presents a schematic plot of the halo picture by sorting all halos in system according to their sizes from the smallest to the largest. Each column in Fig. \ref{fig:1} is a group of halos of the same size. The statistics can be defined on different levels: 1) individual halos; 2) group of halos of the same size (column outlined in Fig. \ref{fig:1}); and 3) global system with all halos of different sizes. 
With the halo picture, we can describe the entire system on four different levels:

First, on the particle level, every dark matter particle, characterized by a mass, $m_{p}$, and a velocity vector, $\boldsymbol{\mathrm{v}}_{\boldsymbol{\mathrm{p}}} $, should belong to one and only one particular (parent) halo. No free particles are allowed in the halo-based description of entire system.

Second, on the halo level, every halo is characterized by a halo size, namely the number of particles in it ($n_{p} $) or, equivalently, the halo mass ($m_{h} $), a one-dimensional (1D) halo virial dispersion ($\sigma _{vh}^{2} $), and halo mean velocity ($\boldsymbol{\mathrm{v}}_{\boldsymbol{\mathrm{h}}} $). The particle velocity $\boldsymbol{\mathrm{v}}_{p} $ can be decomposed into 
\begin{equation} 
\label{ZEqnNum502045} 
\boldsymbol{\mathrm{v}}_{p} =\boldsymbol{\mathrm{v}}_{h} +\boldsymbol{\mathrm{v}}_{p}^{'} ,           
\end{equation} 
namely, the halo mean velocity, $\boldsymbol{\mathrm{v}}_{h} $, and velocity fluctuation, $\boldsymbol{\mathrm{v}}_{p}^{'} $. The halo virial dispersion is defined as
\begin{equation} 
\label{ZEqnNum627513} 
\sigma _{vh}^{2} =var\left(\boldsymbol{\mathrm{v}}_{\boldsymbol{\mathrm{p}}}^{'x} \right)=var\left(\boldsymbol{\mathrm{v}}_{\boldsymbol{\mathrm{p}}}^{'y} \right)=var\left(\boldsymbol{\mathrm{v}}_{\boldsymbol{\mathrm{p}}}^{'z} \right),         
\end{equation} 
that is, the variance of velocity fluctuation of all particles in the same halo. The viral dispersion can be related to the (local) temperature of a halo. The halo mean velocity $\boldsymbol{\mathrm{v}}_{\boldsymbol{\mathrm{h}}} =\langle \boldsymbol{\mathrm{v}}_{\boldsymbol{\mathrm{p}}}^{} \rangle _{h} $ is the mean velocity of all particles in the same halo, where $\left\langle \right\rangle _{h} $ stands for the average over all particles in the same halo.

Third,on the group level (Fig. \ref{fig:1}), the halo group can be characterized by the size of halos in that group ($n_{h} $ or $m_{h} $), halo virial dispersion ($\sigma _{v}^{2} $), and halo velocity dispersion ($\sigma _{h}^{2} $) that is defined as the dispersion (variance) of halo mean velocity, $\boldsymbol{\mathrm{v}}_{\boldsymbol{\mathrm{h}}} $, for all halos in the same group, 
\begin{equation} 
\label{eq:3} 
\sigma _{h}^{2} =var\left(\boldsymbol{\mathrm{v}}_{\boldsymbol{\mathrm{h}}}^{x} \right)=var\left(\boldsymbol{\mathrm{v}}_{\boldsymbol{\mathrm{h}}}^{y} \right)=var\left(\boldsymbol{\mathrm{v}}_{\boldsymbol{\mathrm{h}}}^{z} \right),        
\end{equation} 
which represents the temperature of a halo group due to the motion of halos. The statistics defined on halo group level is the ensemble average for all halos of the same size. The virial dispersion ($\sigma _{v}^{2}$) of a halo group is the average of $\sigma_{vh}^{2} $ for all halos in the same group.

Lastly, on the system level, the entire system can be characterized by the total number of collisionless particles \textit{N} and a 1D velocity dispersion $\sigma _{0}^{2} $ for all \textit{N} particles, which is a measure of the total kinetic energy (or the system temperature) of entire system. The global system statistics can be different from the local statistics in individual halos or halo groups. 

In addition, on the halo level, halos of the same size ($n_{p}$) can have different virial dispersions, $\sigma _{vh}^{2}$, and different mean velocity, $\boldsymbol{\mathrm{v}}_{\boldsymbol{\mathrm{h}}}$. On the group level, a group of halos of the same size ($n_{p}$) have a virial dispersion $\sigma _{v}^{2}=\langle\sigma _{vh}^{2}\rangle_g$. The symbol $\langle \rangle _{g} $ stands for the average over all halos in the same group. Gaussian distribution (Maxwell-Boltzmann) is expected for velocity of all particles in the same halo group (see Fig. \ref{fig:3}). Due to the independence among $\boldsymbol{\mathrm{v}}_{h} $ and $\boldsymbol{\mathrm{v}}_{p}^{'} $ (Eq. \ref{ZEqnNum502045}), 
%\LEt{ No parentheses needed around equation number -- throughout the paper.}
the velocity dispersion in a group can be decomposed into 
\begin{equation} 
\label{ZEqnNum805703} 
\sigma^{2} \left(n_{p} \right)=\sigma _{v}^{2} \left(n_{p} \right)+\sigma _{h}^{2} \left(n_{p} \right),         
\end{equation} 
with two separate contributions, respectively. On the system level, the maximum entropy principle is still valid to describe the statistical equilibrium of the entire system. 

\subsection{Limiting probability distributions}
\label{sec:2.2}
For the problem described, four distributions were identified:\\
1) $X\left(v\right)$: the distribution of 1D velocity \textit{v};\\
2) $Z\left(v\right)$: the distribution of speed (magnitude of velocity);\\
3) $E\left(\varepsilon \right)$: the distribution of particle energy $\varepsilon$;\\
4) $H(\sigma _{v}^{2})$: the distribution of virial dispersion $\sigma _{v}^{2} $, namely, the fraction of particles with a virial dispersion between $\left[\sigma _{v}^{2} ,\sigma _{v}^{2} +d\sigma _{v}^{2} \right]$. The particles' virial dispersion is that of the halo group they belong to. 

%\LEt{ Please integrate points back into paragraph, using 1)..; 2)...; etc.}
\begin{comment}
\begin{enumerate}[leftmargin=\parindent,align=left,labelwidth=\parindent,labelsep=0pt]
\item \noindent $X\left(v\right)$: the distribution of 1D velocity \textit{v};

\item \noindent $Z\left(v\right)$: the distribution of speed (magnitude of velocity);

\item \noindent $E\left(\varepsilon \right)$: the distribution of particle energy $\varepsilon$;

\item \noindent $H(\sigma _{v}^{2})$: the distribution of virial dispersion $\sigma _{v}^{2} $, namely, the fraction of particles with a virial dispersion between $\left[\sigma _{v}^{2} ,\sigma _{v}^{2} +d\sigma _{v}^{2} \right]$. The particles' virial dispersion is that of the halo group they belong to. 
\end{enumerate}
\end{comment}

A relationship between the distributions \textit{X} and \textit{H} can be established through an integral transformation,
\begin{equation}
\label{ZEqnNum517147} 
X\left(v\right)=\int _{0}^{\infty }\frac{1}{\sqrt{2\pi } \sigma } e^{-{v^{2} /2\sigma ^{2} } } H\left(\sigma _{v}^{2} \right)d\sigma _{v}^{2}  ,       
\end{equation} 
where the velocity distribution, $X,$ is expressed as a weighted average of Gaussian distribution of particle velocity in halo group. This average is weighted by the fraction of particles ($H(\sigma _{v}^{2})d\sigma _{v}^{2} $) with a virial dispersion between $[\sigma _{v}^{2} ,\sigma _{v}^{2} +d\sigma _{v}^{2} ]$. The total particle velocity dispersion $\sigma _{}^{2} $ for all particles in the same group is given by Eq. \eqref{ZEqnNum805703}. The \textit{H} distribution is related to the halo mass function and can be obtained by the inverse transform of Eq. \eqref{ZEqnNum517147} \citep{Xu:2021-Mass-functions-of-dark-matter-}.

Similarly, the relation between \textit{Z} and \textit{H} distributions is expressed as:
\begin{equation} 
\label{ZEqnNum453409} 
Z\left(v\right)=\int _{0}^{\infty }\sqrt{\frac{2}{\pi } } \frac{v^{2} }{\sigma ^{3} } e^{-{v^{2} /2\sigma ^{2} } } H\left(\sigma _{v}^{2} \right) d\sigma _{v}^{2}  ,       
\end{equation} 
where the term on the right hand comes from the Maxwellian distribution of particle speed for all particles from the same group. 

\subsection{Virial equilibrium and particle energy}
\label{sec:2.3}
The virial theorem for potential with an exponent of \textit{n} requires 
\begin{equation} 
\label{ZEqnNum790102} 
2\left\langle KE\right\rangle _{g} -n\left\langle PE\right\rangle _{g} =0,         
\end{equation} 
where $\langle KE\rangle _{g} $ and $\langle PE\rangle _{g}$ are particle kinetic and potential energy, respectively, with subscript `\textit{g}' denoting an average over all particles from the same halo group. For the halo group with a total dispersion $\sigma^{2}$, the specific kinetic and potential energy (per unit mass) are: 
\begin{equation}
\left\langle KE\right\rangle _{g} =\left({3/2} \right)\sigma _{}^{2}, \quad \left\langle PE\right\rangle _{g} =\left({2/n} \right)\left\langle KE\right\rangle _{g} =\left({3/n} \right)\sigma _{}^{2}
\label{ZEqnNum282350},
\end{equation}
to satisfy the virial theorem (Eq. \ref{ZEqnNum790102}). Particles in groups of the smallest halos ($m_{h} =0$) have a maximum energy: 
\begin{equation}
\label{ZEqnNum426482} 
\varepsilon _{h} \left(m_{h} =0\right)=\left(\frac{3}{2} +\frac{3}{n} \right)\sigma _{h}^{2} \left(m_{h} =0\right)=\left(\frac{3}{2} +\frac{3}{n} \right)\sigma _{h0}^{2} ,       
\end{equation} 
where $\sigma _{v}^{2} =0$ for the smallest halo group.

From the virial equilibrium (Eq. \ref{ZEqnNum282350}), the average energy, $\varepsilon_v$, for all particles in all halos with the same given speed, $v,$ can be related to the velocity distribution, \textit{X,} as 
\begin{equation} 
\label{eq:18} 
\varepsilon _{v} \left(v\right)dv=2\left(\frac{3}{2}+\frac{3}{n} \right)v^2X\left(v\right)dv,         
\end{equation} 
where the factor of '2' is due to the symmetry with respect to $X(v)=X(-v)$. The energy per particle $\varepsilon \left(v\right)$ with a given speed, $v,$ should be the total energy, $\varepsilon _{v} \left(v\right)dv$, normalized by total number of particles $Z\left(v\right)dv$, namely, the fraction of particles with a speed in $[v,v+dv]$,
\begin{equation} 
\label{ZEqnNum182215} 
\varepsilon \left(v\right)=\frac{\varepsilon _{v} \left(v\right)dv}{Z\left(v\right)dv} =\frac{X\left(v\right)v^{2} }{Z\left(v\right)} \left(3+\frac{6}{n} \right).        
\end{equation} 
This particle energy $\varepsilon \left(v\right)$ is not the instantaneous energy of a particle with given speed $v$. Instead, it is the mean energy of all particles from all halos with a given speed \textit{v}. Finally, a differential equation between \textit{X} and \textit{Z} distributions can be found from Eqs. \eqref{ZEqnNum517147} and \eqref{ZEqnNum453409},
\begin{equation} 
\label{ZEqnNum443368} 
Z\left(v\right)=-2v\frac{\partial X}{\partial v} .          
\end{equation} 
A substitution of Eq. \eqref{ZEqnNum443368} into Eq. \eqref{ZEqnNum182215} gives the particle energy $\varepsilon \left(v\right)$ that is dependent only on the \textit{X} distribution:
\begin{equation}
\label{ZEqnNum273068} 
\varepsilon \left(v\right)=-\frac{X\left(v\right)v}{{\partial X/\partial v} } \left(\frac{3}{2} +\frac{3}{n} \right). 
\end{equation} 

\subsection{Maximum entropy distribution \textbf{X}}
\label{sec:2.4}
The principle of maximum entropy requires velocity distribution with the largest entropy and the least prior information \citep{Jaynes:1957-Information-Theory-and-Statist-I,Jaynes:1957-Information-Theory-and-Statist-II}. On the system level, the principle of maximum entropy is applied to derive the $X$ distribution with two constraints,
\begin{equation} 
\label{ZEqnNum964699} 
\int _{-\infty }^{\infty }X\left(v\right)dv =1,          
\end{equation} 
\begin{equation} 
\label{ZEqnNum377574} 
\int _{-\infty }^{\infty }X\left(v\right)\varepsilon \left(v\right)dv =\left\langle \varepsilon _{1} \right\rangle .         
\end{equation} 
Here, Eq. \eqref{ZEqnNum964699} is the normalization constraint for probability distribution. Equation \eqref{ZEqnNum377574} is an energy constraint requiring the mean particle energy to be a fixed constant $\left\langle \varepsilon _{1} \right\rangle$. The corresponding entropy functional can be constructed as: 
\begin{equation} 
\label{eq:24}
\begin{split}
S\left[X\left(v\right)\right]=&-\int _{-\infty }^{\infty }X\left(v\right)\ln X\left(v\right)dv+\lambda _{1} \left(\int _{-\infty }^{\infty }X\left(v\right)dv- 1\right)\\
&+\lambda _{2} \left(\int _{-\infty }^{\infty }X\left(v\right)\varepsilon \left(v\right)dv- \left\langle \varepsilon _{1} \right\rangle \right),
\end{split}
\end{equation} 
where $\lambda_{1} $ and $\lambda_{2} $ are two Lagrangian multipliers to enforce two constraints. The entropy functional attains its maximum when the functional variation with respect to distribution \textit{X} vanishes,
\begin{equation} 
\label{ZEqnNum605487} 
\frac{\delta S\left(X\left(v\right)\right)}{\delta X} =-\ln X\left(v\right)-1+\lambda _{1} +\lambda _{2} \varepsilon \left(v\right)=0.       
\end{equation} 
The particle energy can be further expressed as (from Eq. \ref{ZEqnNum605487})
\begin{equation} 
\label{ZEqnNum293034} 
\varepsilon \left(v\right)=\frac{1}{\lambda _{2} } \left(\ln X\left(v\right)+1-\lambda _{1} \right).         
\end{equation} 

By equating Eq. \eqref{ZEqnNum293034} with Eq. \eqref{ZEqnNum273068}, a differential equation for distribution $X\left(v\right)$ can be obtained,
\begin{equation} 
\label{eq:27} 
\frac{\partial X}{\partial v} =-\frac{3\lambda _{2} Xv}{2\left(1+\ln \left(X\right)-\lambda _{1} \right)} \left(\frac{2}{n} +1\right).        
\end{equation} 
The general solution of \textit{X} distribution is obtained as:
\begin{equation} 
\label{eq:30} 
X\left(v\right)=\exp \left(\gamma -\sqrt{\alpha ^{2} +\left({v/v_{0} } \right)^{2} } \right).        
\end{equation} 
with parameters $\lambda _{1}$, $\lambda _{2}$ and $n$ equivalently replaced by $\gamma$, $\alpha$, and $v_{0}$. To satisfy the first constraint (Eq. \ref{ZEqnNum964699}), we have:
\begin{equation} 
\label{ZEqnNum854716} 
2e^{\gamma } \alpha v_{0} K_{1} \left(\alpha \right)=1.          
\end{equation} 
Finally, a family of distributions that maximize the system entropy can be obtained for the 1D particle velocity (the \textit{X}-distribution) that depends on two free parameters $\alpha $ and $v_{0} $,  
\begin{equation} 
\label{ZEqnNum864132} 
X\left(v\right)=\frac{1}{2\alpha v_{0} } \frac{e^{-\sqrt{\alpha ^{2} +\left({v/v_{0} } \right)^{2} } } }{K_{1} \left(\alpha \right)} ,         
\end{equation} 
where $K_{y}(x)$ is a modified Bessel function of the second kind,
\begin{equation} 
\label{eq:33} 
K_{x+1} \left(\alpha \right)=K_{x-1} \left(\alpha \right)+\frac{2}{\alpha } xK_{x} \left(\alpha \right).         
\end{equation} 

\section{\textbf{X} distribution for velocity}
\label{sec:3}
\subsection{Statistical properties of the \textbf{X} distribution}
\label{sec:3.1}
In an $X$ distribution (Eq. \ref{ZEqnNum864132}), the parameter $v_{0} $ is introduced as a typical scale of velocity. The shape parameter $\alpha$ dominates the general shape of \textit{X} distribution. The \textit{X} distribution approaches a Laplace (or double-sided exponential) distribution with $\alpha \to 0$ and a Gaussian distribution with $\alpha \to \infty $, respectively. For any intermediate values of $\alpha$, we can approximate the $X$ distribution with a Gaussian distribution   
\begin{equation}
X\left(v\right)=\frac{e^{-\alpha } }{2\alpha v_{0} K_{1} \left(\alpha \right)} \exp \left(-\frac{v^{2} }{2\alpha v_{0}^{2} } \right) \quad   \textrm{for} \quad \left|v\right|\ll v_{0}    
\label{ZEqnNum174426}
,\end{equation}
\noindent and a Laplace distribution (double exponential)
\begin{equation}
X\left(v\right)=\frac{1}{2\alpha v_{0} K_{1} \left(\alpha \right)} \exp \left(-\frac{v}{v_{0} } \right) \quad \textrm{for} \quad \left|v\right|\gg v_{0}.   
\label{ZEqnNum367833}
\end{equation}

\begin{figure}
\includegraphics*[width=\columnwidth]{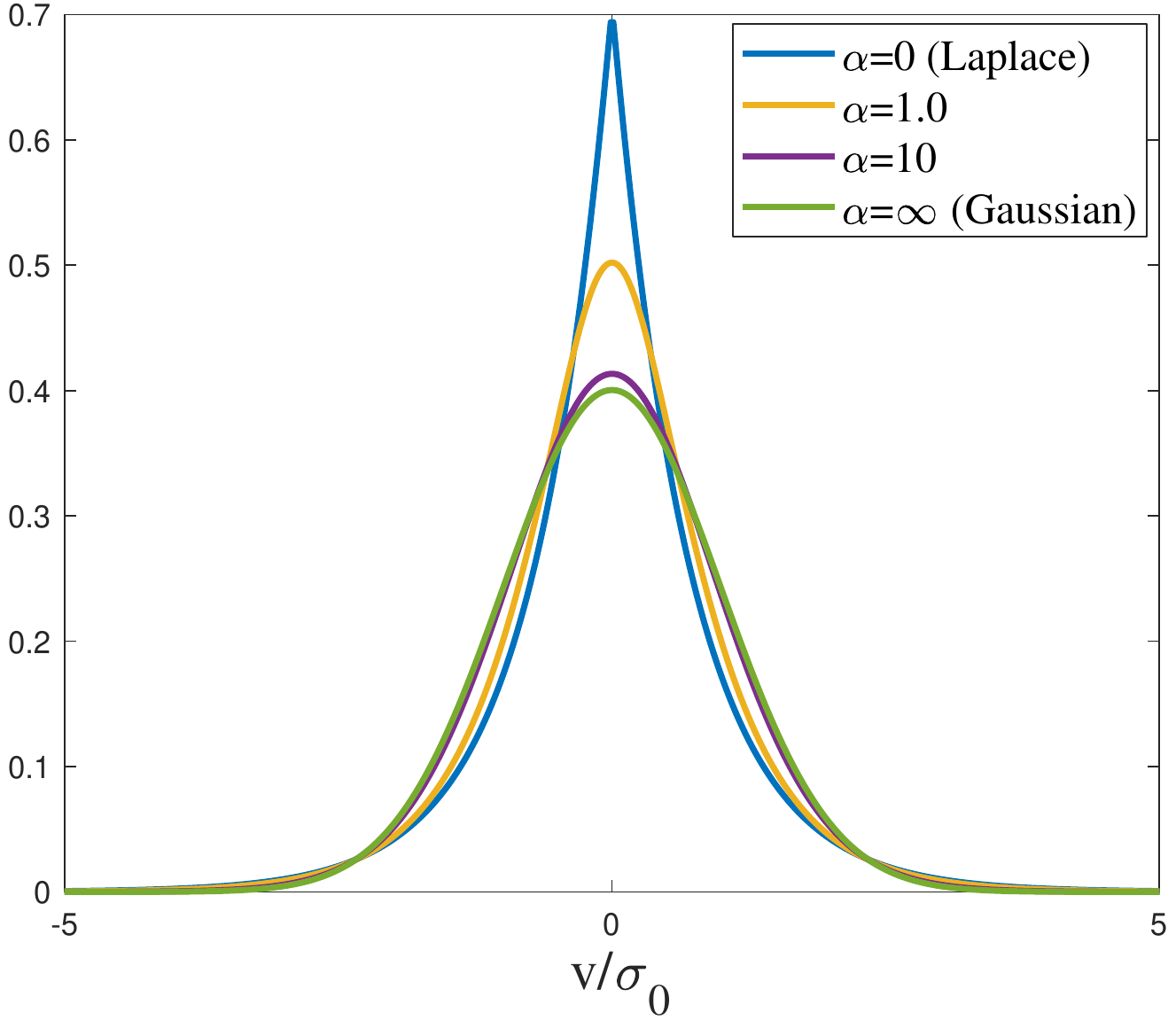}
\caption{\textit{X} distribution with a unit variance for four different shape parameters \textit{$\alpha$}. The \textit{X} distribution approaches a Laplace distribution with $\alpha \to 0$ and a Gaussian distribution with $\alpha \to \infty $, respectively. For intermediate \textit{$\alpha$}, the \textit{X} distribution has a Gaussian core for small velocity $v$ and exponential wings for large \textit{v}, in agreement with N-body simulations}
\label{fig:2}
\end{figure}

The \textit{X} distribution naturally has a Gaussian core for small velocity $v$ (with a variance of $\alpha v_{0}^{2}$) and exponential wings for large velocity values ($v$). This is a remarkable analytical result from maximum entropy principle. Similar features are also observed from cosmological \textit{N}-body simulations \citep{Sheth:2001-Peculiar-velocities-of-galaxie,Cooray:2002-Halo-models-of-large-scale-str}. Figure \ref{fig:2} plots the \textit{X} distribution for four different \textit{$\alpha$}=0, 1, 10, and $\infty $, all with a unit variance (${\langle v^{2}\rangle /\sigma _{0}^{2} } =1$). With decreasing $\alpha$, distribution becomes sharper with a narrower peak and a broader skirt.

The derivation of velocity distribution \textit{X} requires the virial theorem for mechanical equilibrium (Eq. \ref{ZEqnNum182215}), the Maxwellian velocity distribution in halo groups (Eqs. \ref{ZEqnNum517147} and \ref{ZEqnNum453409}), and the maximum entropy principle for statistical equilibrium of global system (Eq. \ref{ZEqnNum605487}). We note that halo mass function is not required, which indicates that the mass function might be related to $X$ distribution as an intrinsic result of maximum entropy. Since the \textit{X} distribution does not explicitly involve parameters characterizing the system ($n$ and $\sigma _{0}^{2} $), additional connections may be identified between $\alpha$, $v_{0}^{2}$, and $n$, $\sigma _{0}^{2}$.

The Gaussian core of velocity distribution \textit{X} mostly comes from particles in small size halos with small viral dispersion $\sigma _{v}^{2} $, where the halo velocity dispersion is much larger than the virial dispersion, namely, $\sigma _{h}^{2} \gg \sigma _{v}^{2} $. On the other side, the exponential wing of velocity distribution is mostly due to the particles in large halos with $\sigma _{v}^{2} \gg \sigma _{h}^{2} $. There is a critical halo mass scale $m_{h}^{*} $, where $\sigma _{h}^{2} (m_{h}^{*} )=\sigma _{v}^{2} (m_{h}^{*} )$. The critical mass, $m_{h}^{*} $, increases with the scale factor \textit{a} (or decreases with the redshift \textit{z}). 

For small halos, the total velocity dispersion $\sigma ^{2}\approx\sigma _{h}^{2} $ with $\sigma _{v}^{2} \approx 0$. Therefore, it is reasonable to assume that the variance of Gaussian core in Eq. \eqref{ZEqnNum174426} is comparable to the halo velocity dispersion, $\sigma _{h0}^{2} $, of small halos, namely:\  
\begin{equation}
\alpha v_{0}^{2} =\sigma ^{2} \approx \sigma _{h0}^{2} \quad,\textrm { \\ where} \quad \sigma _{h0}^{2} =\sigma _{h}^{2} \left(m_{h} \to 0\right).
\label{ZEqnNum643088}
\end{equation}

\noindent We revisit this result in Eqs. \eqref{ZEqnNum426482} and \eqref{ZEqnNum793077}. 

With the limiting velocity distribution explicitly derived in Eq. \eqref{ZEqnNum864132}, its statistical properties can be easily obtained  (listed in Table \ref{tab:A2}). More specifically, the second-order moment (variance) of $X$ distribution should be: 
\begin{equation} 
\label{ZEqnNum108881} 
M_{X} \left(n=2\right)=\alpha \frac{K_{2} \left(\alpha \right)}{K_{1} \left(\alpha \right)} v_{0}^{2} =\sigma _{0}^{2} ,    \end{equation} 
which provides an additional relation between $\alpha $, $v_{0}^{2} $ and $\sigma _{0}^{2}$. 

\subsection{Comparison with N-body simulations}
\label{sec:3.2}
The theory developed here was compared with \textit{N}-body simulations carried out by the Virgo consortium. A comprehensive description of simulation data can be found in \citep{Frenk:2000-Public-Release-of-N-body-simul,Jenkins:1998-Evolution-of-structure-in-cold}. A friends-of-friends algorithm (FOF) was used to identify all halos from simulation data that depends only on a dimensionless parameter \textit{b}, which defines the linking length $b\left({N/V} \right)^{{-1/3} } $, where $V$ is the volume of simulation box. Halos were identified with a linking length parameter of $b=0.2$. All halos were grouped into halo groups according to halo mass $m_{h} $ (or particle number $n_{p}$) (Fig. \ref{fig:1}). 

\begin{figure}
\includegraphics*[width=\columnwidth]{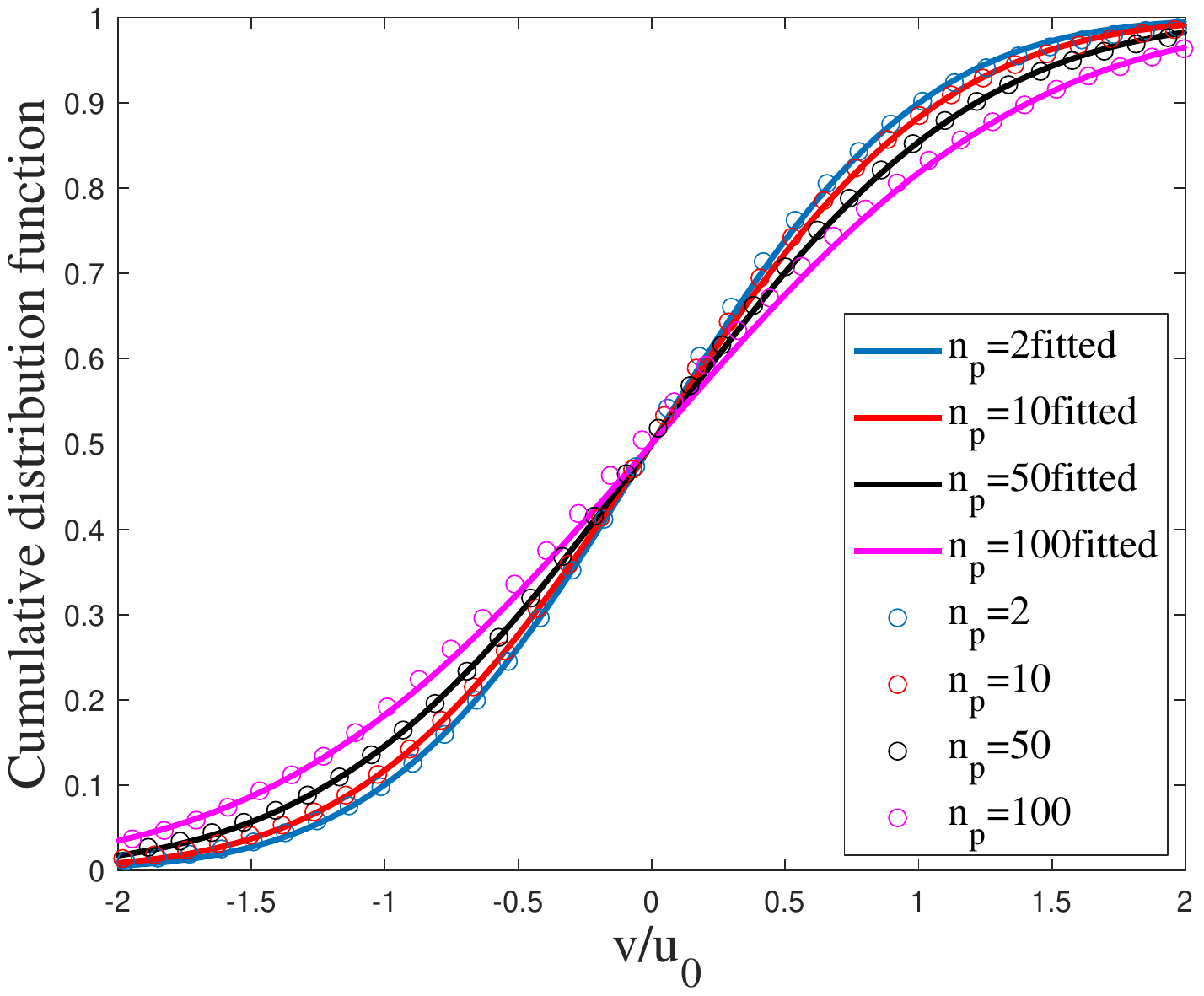}
\caption{Cumulative distribution of particle velocity (normalized by $u_{0}$) in halo groups of different sizes: $n_{h}$=2, 10, 50, and 100. A Gaussian distribution is expected for velocity of all particles in the same group. The cumulative distribution should be error function. Symbols plot the original data from a large-scale \textit{N}-body simulation and lines plot the best fitted error function. Simulation data confirm Gaussian distribution of particle velocity in the same group, while the velocity of all the particles in all halos is non-Gaussian (i.e., the \textit{X} distribution).}
\label{fig:3}
\end{figure}

We first verified the Gaussian distribution of particle velocity in the halo groups  by computing the cumulative distribution function of particle velocity for halo groups of different sizes at a redshift of z=0. For a Gaussian distribution, the cumulative distribution is expected to be an error function. Figure \ref{fig:3} plots the cumulative function from \textit{N}-body simulation (symbols) and the best fit of the simulation data with error functions. The cumulative distribution of particle velocity (normalized by $u_{0} =354.61{km/s} $, i.e. 1D velocity dispersion of entire system at \textit{z}=0) was computed for halo groups of sizes $n_{p} $= 2, 10, 50, and 100. The simulation data confirm a Gaussian distribution of particle velocity for all particles in the same group, regardless of the halo size. However, velocity distribution for all particles (\textit{X} distribution) can be non-Gaussian (Fig. \ref{fig:2}).

\begin{figure}
\includegraphics*[width=\columnwidth]{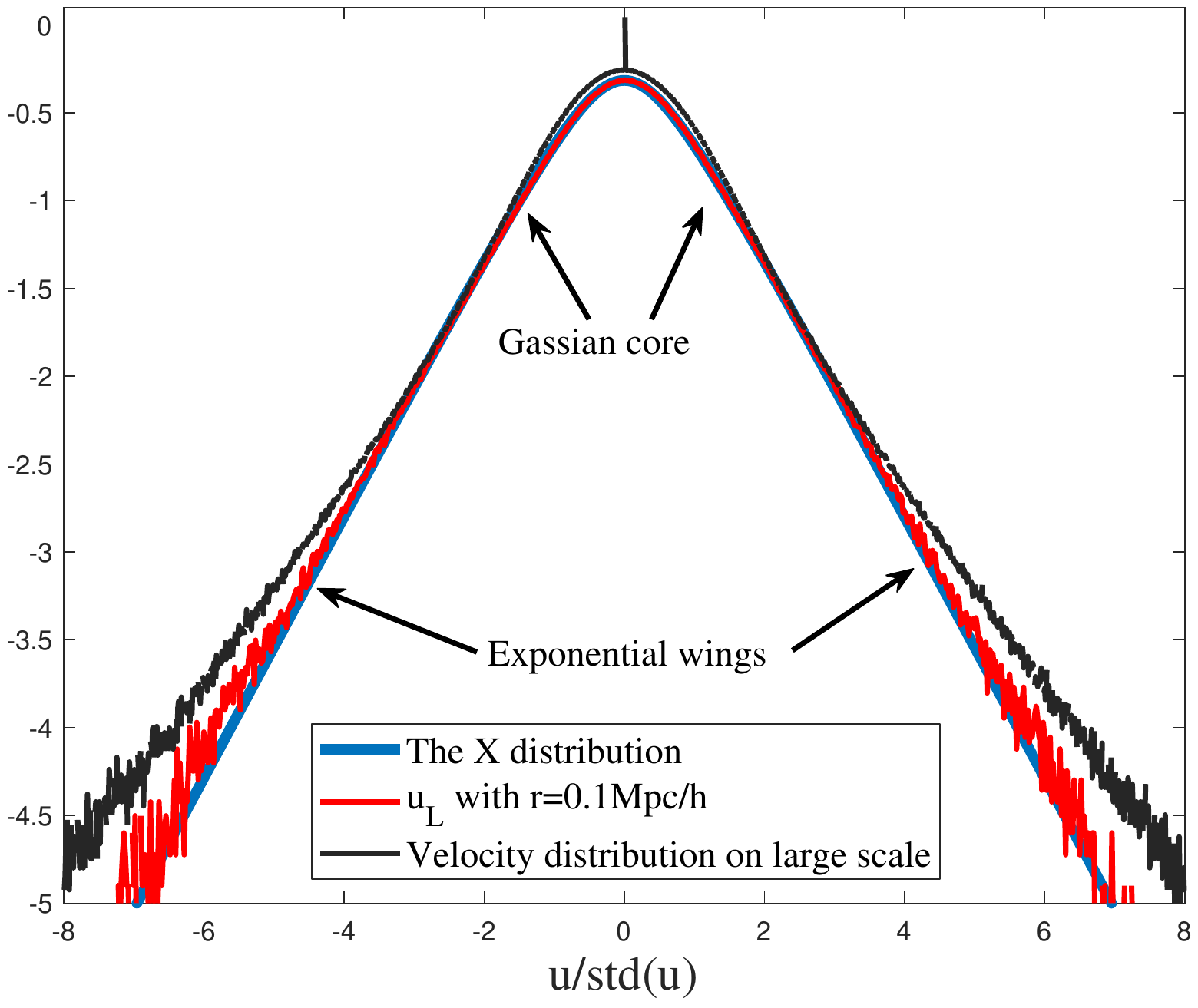}
\caption{ \textit{X} distribution with a unit variance compared against the 1D velocity $u_{L}$ (normalized by std($u_{L}$)) distribution on a small scale, $r,$ from an \textit{N}-body simulation. Vertical axis is plotted in the logarithmic scale (log$_{10}$). The \textit{X} distribution with $\alpha =1.33$ and $v_{0}^{2}={1/3} \sigma_{0}^{2}$ matches the simulated velocity distribution for small separation $r$, where both particle pairs are likely to reside in the same halo and different pairs can be from different halos. The Gaussian core and exponential wings can be clearly identified. For comparison, velocity distribution of all DM particles from the same simulation is also presented (black line). System on large scale is not fully virialized, which leads to the deviation from the predicted $X$ distribution.}
\label{fig:4}
\end{figure}

Figure \ref{fig:4} presents a comparison of $X$ distribution (Eq. \ref{ZEqnNum864132}) and simulation data for the 1D velocity on a fixed small-scale value of \textit{r}. We first identify pairs of particles with a given separation r = 0.1 Mpc/h at \textit{z}=0. These pairs of particles likely reside in the same halo because of the small separation $r$. The 1D velocity, $u_{L}$, was then computed as the projection of particle velocity, $\boldsymbol{\mathrm{u,}}$ along the direction of separation $\boldsymbol{\mathrm{r}}$, namely, $u_{L} =\boldsymbol{\mathrm{u}}\cdot \boldsymbol{\mathrm{r}}$. The small-scale particle velocity, $u_{L}$, is finally normalized to have a unit variance and compared with the \textit{X} distribution. Figure \ref{fig:4} demonstrates the good agreement between the simulated distribution of DM velocity on a small scale (longitudinal velocity $u_L$) and the predicted $X$ distribution. The noise at large velocity mostly comes from insufficient number of samples for large $u_L$. The best fit to simulation leads to parameters $\alpha =1.33$ and $v_{0}^{2} ={1/3} \sigma _{0}^{2} $, where $\sigma _{0}^{2} =var\left(u_{L} \right)$. The scale and redshift dependence of velocity distributions were also systematically studied in an separate paper \citep{Xu:2022-Two-thirds-law-for-pairwise-ve}. With kinetic energy cascaded to larger scales, the symmetric velocity distribution on a small scale (Fig. \ref{fig:4}) gradually deviates from $X$ distribution and becomes asymmetric with increasing skewness \citep{Xu:2022-Two-thirds-law-for-pairwise-ve}.  

In N-body simulations, small halos tend to be more virialized. The core region of large halos also tends to be more virialized than the outskirt region. Large halos are not fully virialized and the velocity distribution might not be exactly Maxwellian. Therefore, system on large scale is not fully virialized, which leads to the deviation of velocity distribution from the predicted $X$ distribution. To illustrate this, we also present the velocity distribution of all DM particles from the same N-body simulation (black line in Fig. \ref{fig:4}), where the deviation can be observed for velocity distribution on large scale. That deviation is mostly for large velocity in large halos and the outskirts region. The good agreement still holds for low velocity values.   

\section{\textbf{Z} and \textbf{E} distributions for speed and energy}
\label{sec:4}
\begin{figure}
\includegraphics*[width=\columnwidth]{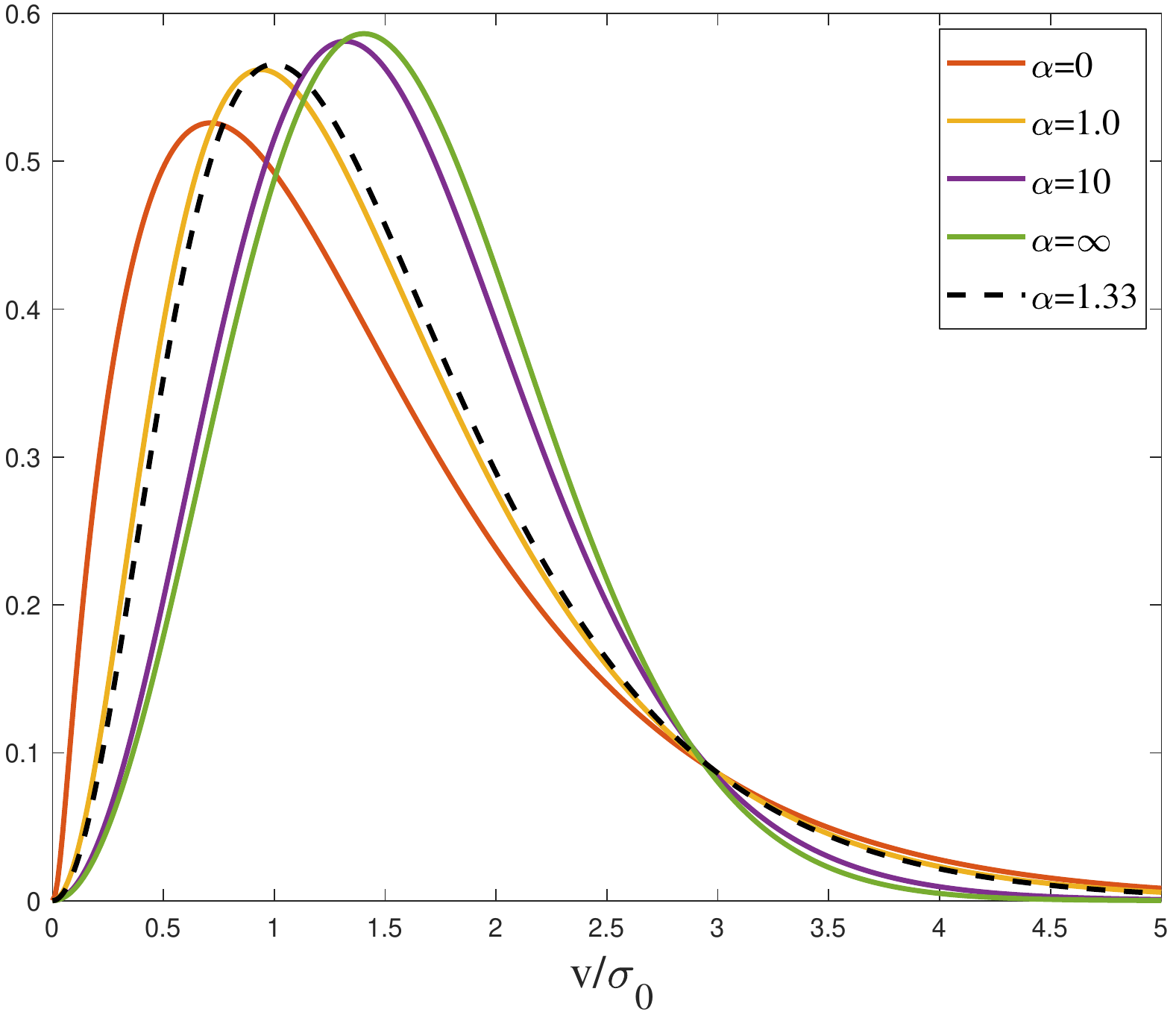}
\caption{Distribution of particle speed (\textit{Z} distribution) for different $\alpha $ ($\alpha $\textit{ }= 0.1, 1, 10, $\mathrm{\infty}$). The \textit{Z} distribution approaches a Maxwell-Boltzmann distribution with $\alpha \to \infty $. With increasing \textit{$\alpha$}, the distribution shifts towards the large velocity with more particles exhibiting an intermediate speed and, conversely, fewer particles exhibiting a small and large speed.}
\label{fig:5}
\end{figure}

With the velocity distribution (\textit{X} distribution) explicitly derived in Eq. \eqref{ZEqnNum864132}, the distribution of particle speed for entire system (\textit{Z} distribution) can be obtained from Eq. \eqref{ZEqnNum443368},
\begin{equation} 
\label{ZEqnNum762113} 
Z\left(v\right)=\frac{1}{\alpha K_{1} \left(\alpha \right)} \cdot \frac{v^{2} }{v_{0}^{3} } \cdot \frac{e^{-\sqrt{\alpha ^{2} +\left({v/v_{0} } \right)^{2} } } }{\sqrt{\alpha ^{2} +\left({v/v_{0} } \right)^{2} } } .        
\end{equation} 
Figure \ref{fig:5} plots the particle speed distribution (\textit{Z} distribution) for different $\alpha $ ($\alpha $= 0, 1, 10, $\mathrm{\infty}$) with $\sigma _{0}^{2} =1$. The \textit{Z} distribution approaches a Maxwell-Boltzmann distribution with $\alpha \to \infty $. Specifically, with increasing $\alpha $, \textit{Z} distribution shifts towards large velocity with more particles have an intermediate speed. Statistical properties of \textit{Z} distribution are also listed in Table \ref{tab:A2}. 

Finally, particle energy $\varepsilon \left(v\right)$ reads (Eqs. \ref{ZEqnNum182215}, \ref{ZEqnNum864132}, and \ref{ZEqnNum762113})
%\LEt{ Parentheses around equation numbers are not necessary.}
\begin{equation} 
\label{ZEqnNum520740} 
\varepsilon \left(v\right)=\frac{3}{2} \left(1+\frac{2}{n} \right)v_{0}^{2} \sqrt{\alpha ^{2} +\left(\frac{v}{v_{0} } \right)^{2} } .        
\end{equation} 
In the kinetic theory of gases, the energy of molecules is of a kinetic nature and proportional to $v^{2}$, i.e. $\varepsilon(v)=3v^2/2$. While for collisionless dark matter particles, mean particle energy (including both kinetic and potential) for all particles with the same speed $v$ follows a parabolic scaling when $v\ll v_{0} $ and a linear scaling when $v\gg v_{0} $,
\begin{equation}
\varepsilon \left(v\right)\approx \frac{3}{2} \left(1+\frac{2}{n} \right)\left(\alpha v_{0}^{2} +\frac{v^{2} }{2\alpha } \right) \quad \textrm{for} \quad v\ll v_{0},     
\label{ZEqnNum265529}
\end{equation}
\noindent and
\begin{equation}
\varepsilon \left(v\right)\approx \frac{3}{2} \left(1+\frac{2}{n} \right)v_{0}^{} v \quad \textrm{for} \quad v\gg v_{0}.
\label{ZEqnNum798047}
\end{equation}

\noindent This unique scaling might be critical to understand the "deep-MOND" behavior in  Modified Newtonian Dynamics (MOND) theory \citep{Xu:2022-The-origin-of-MOND-acceleratio}. Particles in the outer region of halos with $v\gg v_{0}$ are more influenced by the long range gravity from other halos, which leads to the linear scaling of mean particle energy and "non-Newtonian" behavior. By contrast, particles in the inner region are less influenced and exhibit a "Newtonian" behavior.   

\begin{figure}
\includegraphics*[width=\columnwidth]{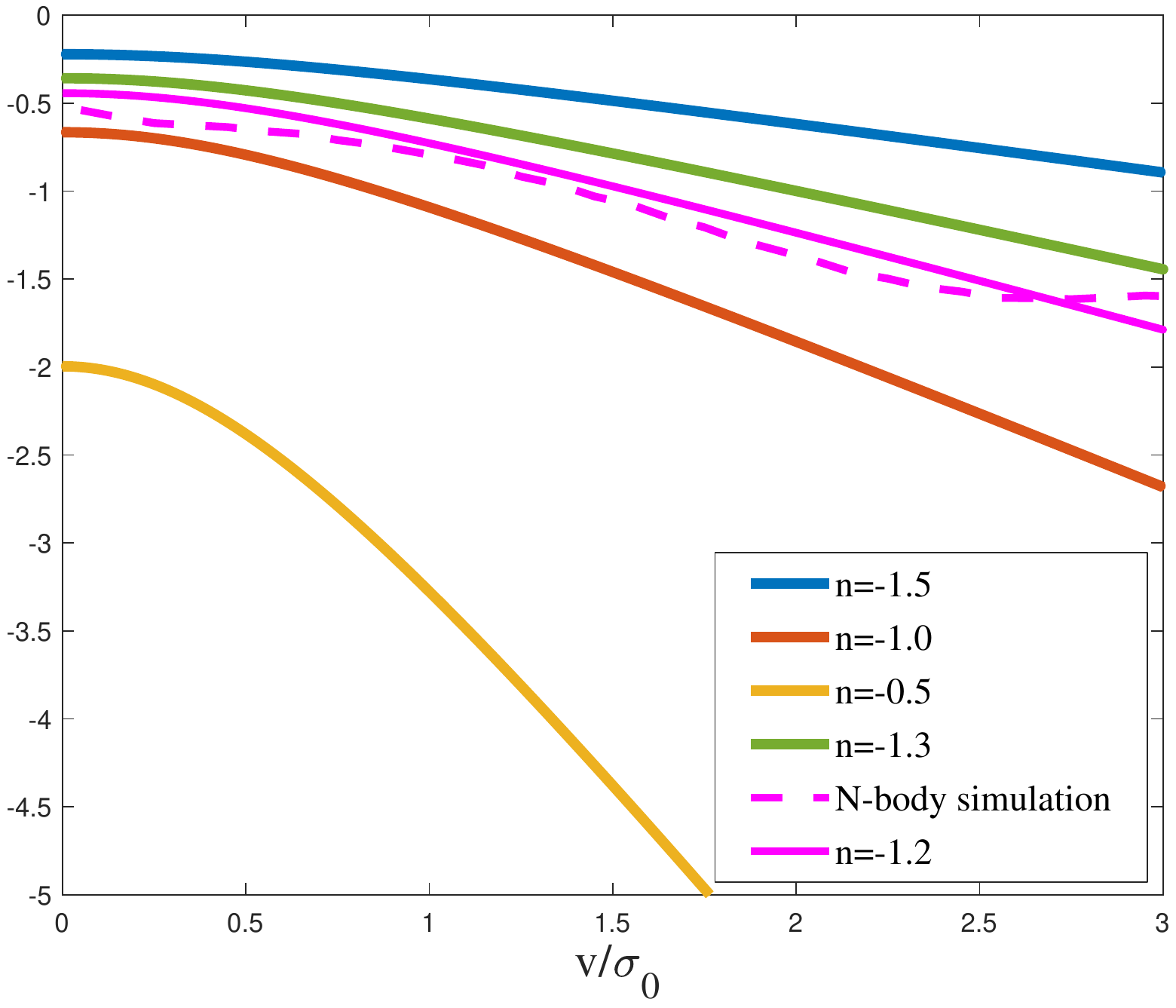}
\caption{Dependence of normalized particle energy $\varepsilon \left(v\right)$ (normalized by $\sigma_0^2$) on particle speed \textit{v} for five different potential exponents \textit{n} with a fixed shape parameter $\alpha =1.33$. For small speed, the particle energy follows a parabolic law with particle speed ($\varepsilon \left(v\right)\propto v^{2} $ Eq. \eqref{ZEqnNum265529}). While for large speed, the particle energy follows a linear scaling with particle speed ($\varepsilon \left(v\right)\propto v$ in Eq. \ref{ZEqnNum798047}) that is different from gas molecules. Particle energy from N-body simulation is plotted as the dash line. Simulation matches an effective exponent $n=-1.2$ for virial theorem (not -1 due to the mass cascade and nonzero halo surface energy \citep{Xu:2021-Inverse-mass-cascade-mass-function}}
\label{fig:6}
\end{figure}

Figure \ref{fig:6} shows the plot of the dependence of normalized mean particle energy $\varepsilon(v)$ on particle speed, $v,$ for five different potential exponents \textit{n} with a fixed $\alpha =1.33$. Both parabolic and linear scaling are clearly shown in Fig. \ref{fig:6} for small and large speeds, respectively. Dash line presents mean particle energy from the same \textit{N}-body simulation. Velocity is normalized by the 1D velocity dispersion for all particles in all halos ($\sigma _{0} =395.18{km/s}$). The average total energy for all particles in all halos with a given speed \textit{v} is computed for each speed, \textit{v}. The deviation at large velocity might be due to the insufficient sampling. Simulation matches an effective potential exponent $n\approx-1.2\ne-1$ for virial theorem in Eq. \eqref{ZEqnNum282350}.

\begin{figure}
\includegraphics*[width=\columnwidth]{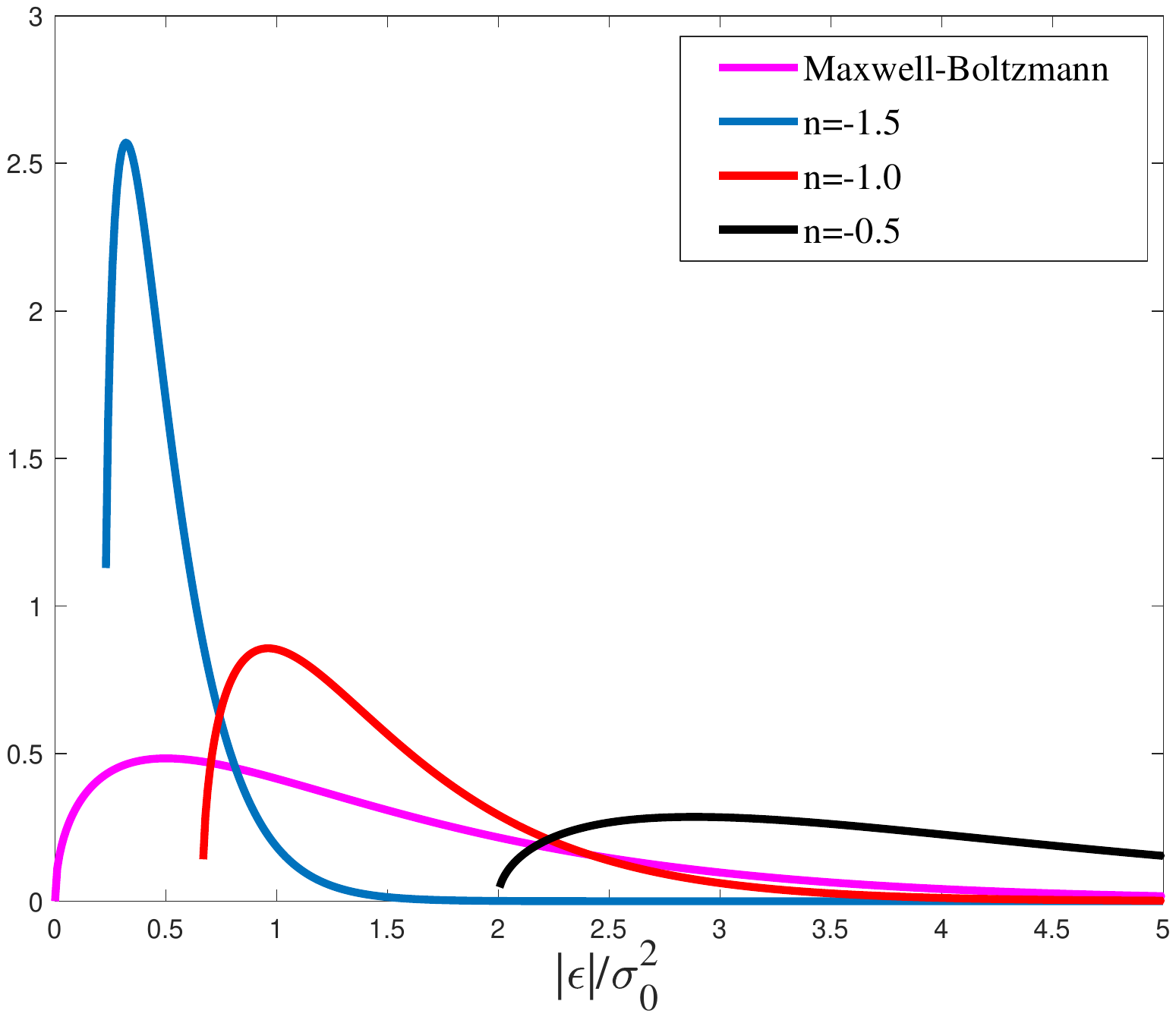}
\caption{Particle energy distribution, $E,$ for three different potential exponents \textit{n} = -1.5, -1.0, and -0.5 with a fixed $\alpha =1.33$. For comparison, the energy distribution of a Maxwell-Boltzmann velocity statistics is also presented in the same plot.}
\label{fig:7}
\end{figure}

The mean particle energy of all particles reads
\begin{equation} 
\label{ZEqnNum371523} 
\left\langle \varepsilon \right\rangle =\left\langle \varepsilon _{h} \right\rangle =\int _{0}^{\infty }Z\left(v\right) \varepsilon \left(v\right)dv=\left(\frac{3}{2} +\frac{3}{n} \right)\sigma _{0}^{2} .       
\end{equation} 
The particle energy distribution (\textit{E} distribution) can be found as (with $Z\left(v\right)$ from Eq. \eqref{ZEqnNum762113} and $\varepsilon \left(v\right)$ from Eq. \eqref{ZEqnNum520740}):
\begin{equation} 
\label{ZEqnNum517959} 
E\left(\varepsilon \right)=\frac{Z\left(v\right)}{{d\varepsilon /dv} } =-\frac{2n}{3\left(n+2\right)} \frac{e^{-\gamma } \sqrt{\gamma ^{2} -\alpha ^{2} } }{\alpha K_{1} \left(\alpha \right)v_{0}^{2} } ,       
\end{equation} 
where the dimensionless particle energy $\gamma $ is defined as 
\begin{equation}
\label{ZEqnNum671554} 
\gamma =\frac{2n}{3\left(n+2\right)} \frac{\varepsilon }{v_{0}^{2} } .          
\end{equation} 
With $\gamma \ge \alpha $, there exists a maximum particle energy (or minimum in absolute value) from Eq. \eqref{ZEqnNum671554} corresponding to particles in the smallest halo groups (Eq. \ref{ZEqnNum426482}), where
\begin{equation} 
\label{ZEqnNum793077} 
\varepsilon _{\max } =\frac{3}{2} \left(1+\frac{2}{n} \right)\alpha v_{0}^{2} .          
\end{equation} 
For comparison, the energy distribution for Maxwell-Boltzmann velocity distribution is:
\begin{equation} 
\label{eq:56} 
f_{MB} \left(\varepsilon \right)=2\sqrt{\frac{\varepsilon }{\pi \sigma _{0}^{2} } } \frac{1}{\sigma _{0}^{2} } e^{{-\varepsilon /\sigma _{0}^{2} } } .         
\end{equation} 
Figure \ref{fig:7} plots the energy distribution for three different potential exponents n=-1.5, -1.0, and -0.5 with a fixed $\alpha =1.33$. Compared to the Maxwell-Boltzmann, more dark matter particles have low energy and less particles have high energy.  

In principle, both halo virial dispersion ($\sigma _{v}^{2}$ for temperature of halos) and halo velocity dispersion ($\sigma _{h}^{2}$ for temperature of halo groups) are functions of halo size $n_p$ or $m_h$. However, they can scale very differently with the halo size, where $\sigma _{h}^{2} \ll \sigma _{v}^{2} $ for massive and hot halos and $\sigma _{v}^{2} \ll \sigma _{h}^{2} $ for small halos. The halo virial dispersion scales with the halo size as $\sigma _{v}^{2} \propto m_{h}^{{1/\beta}}$, where $\beta ={1/\left(1+{n/3} \right)}$. 

The maximum entropy distributions we derived have two free parameters $\alpha$ and $v_{0}$, while the system is fully characterized by the potential exponent, \textit{n}, particle velocity dispersion, $\sigma _{0}^{2} $, and the halo velocity dispersion, $\langle \sigma _{h}^{2} \rangle $. Equation \eqref{ZEqnNum108881} provides a connection with $\sigma _{0}^{2} $. Another connection can be found by identifying the maximum particle energy, $\varepsilon(v),$ at $v=0$ in Eq. \eqref{ZEqnNum793077},  which is the mean energy of all particles with a vanishing speed in all halos. The maximum particle energy, $\varepsilon _{h}(\sigma _{v}^{2} ),$ among particles in all halos is $({3/2} +{3/n} )\langle \sigma _{h}^{2} \rangle, $ with $\sigma _{v}^{2} =0$ for particles in the smallest halos from Eq. \eqref{ZEqnNum426482}. Since most particles with small speed reside in small halos with $\sigma _{v}^{2} =0$, 
\begin{equation}
\alpha v_{0}^{2} \approx \left\langle \sigma _{h}^{2} \right\rangle,
\label{eq:57}
\end{equation}

\noindent which is the same as we discussed in Eq. \eqref{ZEqnNum643088}. With the help from Eq. \eqref{ZEqnNum108881}  (the variance of \textit{X} distribution), we can easily write:
\begin{equation} 
\label{ZEqnNum225364} 
\frac{\left\langle \sigma _{h}^{2} \right\rangle }{\sigma _{0}^{2} } =\frac{K_{1} \left(\alpha \right)}{K_{2} \left(\alpha \right)} ,          
\end{equation} 
from which two extremes can be confirmed, that is, $\langle \sigma _{h}^{2} \rangle =0$ for $\alpha\to0$ and $\langle \sigma _{h}^{2} \rangle =\sigma _{0}^{2} $ for $\alpha\to\infty$ (Table \ref{tab:A1}).

Non-bonded interactions can be generally classified into two categories: short-range and long- range interactions \citep{Cheung:2002-Structures-and-properties-of-l}. A force is defined to be long-range if it decreases with the distance slower than $r^{-d} $, where \textit{d} is the dimension of the system. Therefore, the pair interaction potential is long-range for $n>-2$ and short-range for $n<-2$ in a 3D space with \textit{d}=3. For short-range force with $n<-2$, we expect the system is not halo-based with $\alpha \to \infty $ and \textit{X} distribution approaches a Gaussian for $n=-2$. For long-range force with $n>-2$, a halo-based system is expected in order to maximize system entropy. With $\alpha \to 0$, the \textit{X} distribution approaches a Laplace distribution for $n\to 0$. The shape parameter $\alpha $ reflects the nature of force (short or long range) and should be related to the potential exponent \textit{n}. 

\section{Conclusions}
\label{sec:5}
The maximum entropy distributions of dark matter velocity, speed, and energy are analytically derived. They are the limiting distributions on small scales. The formation for halo structures is a direct result of entropy maximizing for systems involving long-range interactions. The virial theorem is applied for mechanical equilibrium, while the maximum entropy principle is applied for the statistical equilibrium on system level. The predicted maximum entropy distribution of velocity in entire system (\textit{X} in Eq. \ref{ZEqnNum864132}) naturally exhibits a Gaussian core at small velocity and exponential wings at large velocity. Prediction is compared with a \textit{N}-body simulation with good agreement (Fig. \ref{fig:4}). The speed (\textit{Z} in Eq. \ref{ZEqnNum762113} and Fig. \ref{fig:5}) and energy (\textit{E} in Eq. \ref{ZEqnNum517959} and Fig. \ref{fig:7}) distributions are also presented. 

The standard kinetic energy is proportional to $v^{2}$ in kinetic theory of gases with short range forces. Dark matter particles in halo-based $\Lambda$CDM cosmology with long-range interactions have a mean total energy that follows a parabolic scaling when $v\ll v_{0}$ and a linear scaling when $v\gg v_{0} $ (Eq. \ref{ZEqnNum520740} and Fig. \ref{fig:6}), where $v_{0}$ is a typical velocity scale. The shape parameter $\alpha$ reflects the nature of force (long or short range) and should be related to the potential exponent \textit{n}. For systems involving short-range interactions, the Gaussian is the maximum entropy distribution and no halo structures are formed. For systems with long-range interactions, we find that substructures (halos and halo groups) are required to form to maximize system entropy. Though velocity in substructure is still Gaussian, velocity in  entire system can be non-Gaussian and follows a more general distribution (\textit{X} distribution). Since the particle velocity must follow \textit{X} distribution to maximize entropy, a broad spectrum of halos with different sizes must be formed as a direct result of entropy maximizing via mass and energy cascade. The halo mass function is given by the \textit{H} distribution that is related to the \textit{X} distribution via Eq. \eqref{ZEqnNum517147}. In this regard, halo mass function is an intrinsic distribution to maximize the system entropy \citep{Xu:2021-Mass-functions-of-dark-matter-}.

\section*{Acknowledgements}
This research was supported by Laboratory Directed Research and Development at Pacific Northwest National Laboratory (PNNL). PNNL is a multiprogram national laboratory operated for the U.S. Department of Energy (DOE) by Battelle Memorial Institute under Contract no. DE-AC05-76RL01830. Two datasets underlying this article, i.e. a halo-based and correlation-based statistics of dark matter flow, are available on Zenodo \citep{Xu:2022-Dark_matter-flow-dataset-part1,Xu:2022-Dark_matter-flow-dataset-part2}, along with the accompanying presentation slides "A comparative study of dark matter flow \& hydrodynamic turbulence and its applications" \citep{Xu:2022-Dark_matter-flow-and-hydrodynamic-turbulence-presentation}.

\bibliographystyle{Papers}
\bibliography{Papers}

%\appendix
%\section{Statistical properties of \textbf{X} and \textbf{Z} distributions}
%Please see Table \ref{tab:A2}.

\begin{table*}
\caption{\textit{X} distribution family and parameters for different potential exponents \textit{n}}
\begin{tabular}{p{0.4in}p{0.4in}p{0.8in}p{0.8in}p{0.8in}p{0.8in}p{0.8in}p{0.8in}} 
\hline 
$n$ & $\beta $ & $\alpha $ & $v_{0}^{2} $ & $\left\langle \sigma _{h}^{2} \right\rangle $ & $\left\langle \sigma _{v}^{2} \right\rangle $ & $X\left(v\right)$ & Distribution \\ 
\hline 
0 & 1 & 0 & $\frac{\sigma _{0}^{2}}{2} $ & 0 & $\sigma _{0}^{2} $ & $\frac{e^{{-\sqrt{2} v/\sigma _{0}^{} } } }{\sqrt{2} \sigma _{0}^{} } $ & Laplace \\ 
\hline 
-1 & $\frac{3}{2} $& $\frac{K_{1} \left(\alpha \right)}{K_{2} \left(\alpha \right)} =\frac{\left\langle \sigma _{h}^{2} \right\rangle }{\sigma _{0}^{2} } $ & $\frac{\sigma _{0}^{2} K_{1} \left(\alpha \right)}{\alpha K_{2} \left(\alpha \right)} $ & $\frac{\sigma _{0}^{2} K_{1} \left(\alpha \right)}{K_{2} \left(\alpha \right)} $ & $\left[1-\frac{K_{1} \left(\alpha \right)}{K_{2} \left(\alpha \right)} \right]\sigma _{0}^{2} $ & $\frac{e^{-\sqrt{\alpha ^{2} +\left({v/v_{0} } \right)^{2} } } }{2\alpha v_{0} K_{1} \left(\alpha \right)} $ & \textit{X} distribution \\ 
\hline 
-2 & 3 & $\infty $ & 0 & $\sigma _{0}^{2} $ & 0 & $\frac{e^{{-v^{2} /2\sigma _{0}^{2} } } }{\sqrt{2\pi } \sigma _{0}^{} } $ & Gaussian \\ \hline 
\end{tabular}
\label{tab:A1}
\end{table*}

\begin{table*}
\caption{Statistical properties of \textit{X} and \textit{Z} distributions}
\begin{tabular}{p{1.5in}p{2.2in}p{2.8in}} \hline 
Distribution name & X & Z \\ \hline 
Support  & $(-\infty ,+\infty )$ & $[0,+\infty )$ \\ \hline 
PDF & $\frac{1}{2\alpha v_{0} } \frac{e^{-\sqrt{\alpha ^{2} +\left({x/v_{0} } \right)^{2} } } }{K_{1} \left(\alpha \right)} $ & $\frac{1}{\alpha K_{1} \left(\alpha \right)} \cdot \frac{x^{2} }{v_{0}^{3} } \cdot \frac{e^{-\sqrt{\alpha ^{2} +\left({x/v_{0} } \right)^{2} } } }{\sqrt{\alpha ^{2} +\left({x/v_{0} } \right)^{2} } } $ \\ \hline 
CDF & $\frac{1}{2} \left(1+\frac{J_{a} \left(\sqrt{\alpha ^{2} +\left({x/v_{0} } \right)^{2} } \right)+\left|{x/v_{0} } \right|e^{-\sqrt{\alpha ^{2} +\left({x/v_{0} } \right)^{2} } } }{Sign(x)\alpha K_{1} \left(\alpha \right)} \right)$ & $\frac{J_{\alpha } \left(\sqrt{\alpha ^{2} +\left({x/v_{0} } \right)^{2} } \right)}{\alpha K_{1} \left(\alpha \right)} $ \\ \hline 
Mean & 0 & $\sqrt{\frac{8\alpha }{\pi } } \frac{K_{{3/2} } \left(\alpha \right)}{K_{1} \left(\alpha \right)} v_{0}^{} $  \\ \hline 
Variance & $\alpha \frac{K_{2} \left(\alpha \right)}{K_{1} \left(\alpha \right)} v_{0}^{2} $ & $\left[3\alpha \frac{K_{2} \left(\alpha \right)^{} }{K_{1} \left(\alpha \right)^{} } -\frac{8\alpha }{\pi } \frac{K_{{3/2} } \left(\alpha \right)^{2} }{K_{1} \left(\alpha \right)^{2} } \right]v_{0}^{2} $ \\ \hline 
Second moment & $\alpha \frac{K_{2} \left(\alpha \right)}{K_{1} \left(\beta \right)} v_{0}^{2} $ & $3\alpha \frac{K_{2} \left(\alpha \right)}{K_{1} \left(\alpha \right)} v_{0}^{2} $  \\ \hline 
Moments & $\frac{\left(2\alpha \right)^{{m/2} } \Gamma \left({\left(1+m\right)/2} \right)}{\sqrt{\pi } } \cdot \frac{K_{\left(1+{m/2} \right)} \left(\alpha \right)}{K_{1} \left(\alpha \right)} v_{0}^{m} $  & $\frac{\left(2\alpha \right)^{{1+m/2} } \Gamma \left({\left(3+m\right)/2} \right)}{\sqrt{\pi } } \cdot \frac{K_{\left(1+{m/2} \right)} \left(\alpha \right)}{aK_{1} \left(\alpha \right)} v_{0}^{m} $  \\ \hline 
Generalized kurtosis & $\left(\frac{2K_{1} \left(\alpha \right)}{K_{2} \left(\alpha \right)} \right)^{{m/2} } \frac{\Gamma \left({\left(1+m\right)/2} \right)}{\sqrt{\pi } } \cdot \frac{K_{\left(1+{m/2} \right)} \left(\alpha \right)}{K_{1} \left(\alpha \right)} $ & $\left(\frac{2K_{1} \left(\alpha \right)}{3K_{2} \left(\alpha \right)} \right)^{{m/2} } \frac{2\Gamma \left({\left(3+m\right)/2} \right)}{\sqrt{\pi } } \cdot \frac{K_{\left(1+{m/2} \right)} \left(\alpha \right)}{K_{1} \left(\alpha \right)} $ \\ \hline 
Entropy & $1+\alpha \frac{K_{0} \left(\alpha \right)}{K_{1} \left(\alpha \right)} +\ln \left(2\alpha v_{0} K_{1} \left(\alpha \right)\right)$  & $\gamma +\ln \left(2v_{0} K_{1} \left(\alpha \right)\right)+\frac{K_{0} \left(\alpha \right)}{K_{1} \left(\alpha \right)} \frac{\left(\alpha ^{2} -1\right)}{\alpha } +\frac{\int _{\alpha }^{\infty }e^{-t} \sqrt{t^{2} -\alpha ^{2} } \ln \left(t\right)dt }{\alpha K_{1} \left(\alpha \right)} $ \\ \hline 
Moment-generating function & $\frac{K_{1} \left(\alpha \sqrt{1-\left(v_{0} t\right)^{2} } \right)}{K_{1} \left(\alpha \right)\sqrt{1-\left(v_{0} t\right)^{2} } } $ & $\frac{2}{\sqrt{\pi } } \sum _{m=0}^{\infty }\Gamma \left(\frac{3+m}{2} \right)\frac{\left(\sqrt{2\alpha } v_{0} t\right)^{m} }{m!}  \frac{K_{\left(1+{m/2} \right)} \left(\alpha \right)}{K_{1} \left(\alpha \right)} $   or $MGF_{Z} \left(t\right)$  \\ \hline 
Characteristic function & $\frac{K_{1} \left(\alpha \sqrt{1+\left(v_{0} t\right)^{2} } \right)}{K_{1} \left(\alpha \right)\sqrt{1+\left(v_{0} t\right)^{2} } } $ & $\frac{2}{\sqrt{\pi } } \sum _{m=0}^{\infty }\Gamma \left(\frac{3+m}{2} \right)\frac{\left(\sqrt{2\alpha } v_{0} it\right)^{m} }{m!}  \frac{K_{\left(1+{m/2} \right)} \left(\alpha \right)}{K_{1} \left(\alpha \right)} $ \\ \hline 
Maximum entropy constraint & $E\left(\sqrt{\alpha ^{2} +\left(\frac{x}{v_{0} } \right)^{2} } \right)=1+\alpha \frac{K_{0} \left(\alpha \right)}{K_{1} \left(\alpha \right)} $ &  \\ \hline 
\end{tabular}
\label{tab:A2}

\begin{equation}
\begin{split}
&\textrm{where} \quad J_{s} \left(x\right) \textrm{function is defined as the integral:}\\
&J_{s} \left(x\right)=\int _{s}^{x}e^{-t} \sqrt{t^{2} -s^{2} }  dt,\quad J_{s} \left(s\right)=0 \quad \textrm{and} \quad J_{s} \left(\infty \right)=sK_{1} \left(s\right)
\end{split}
\end{equation}
\begin{equation}
P\left(t\right)=\int _{0}^{\infty }X\left(x\right) e^{xt} dx=\frac{v_{0} te^{-\alpha } \left(1+\alpha \right)+J_{\alpha \sqrt{1-\left(v_{0} t\right)^{2} } } \left(\alpha \right)}{2\alpha K_{1} \left(\alpha \right)\left[1-\left(v_{0} t\right)^{2} \right]} +\frac{K_{1} \left(\alpha \sqrt{1-\left(v_{0} t\right)^{2} } \right)}{2K_{1} \left(\alpha \right)\sqrt{1-\left(v_{0} t\right)^{2} } } 
\end{equation}
\begin{equation}
\begin{split}
&MGF_{Z} \left(t\right)=\int _{0}^{\infty }Z\left(x\right) e^{xt} dx=2\left[P\left(t\right)+t\frac{\partial P}{\partial t} \right]\\ &\textrm{Euler constant} \quad \gamma \approx 0.5772
\end{split}
\end{equation}
\end{table*}
\label{lastpage}
\end{document}